\documentstyle[epsf]{article}

\catcode`@=11
\def\[{\relax\ifmmode\@badmath\else
 \begin{trivlist}%
 \@beginparpenalty\predisplaypenalty
 \@endparpenalty\postdisplaypenalty
 \item[]\leavevmode
 \hbox to\linewidth\bgroup $\m@th\displaystyle
 \hskip\mathindent\bgroup\fi}

\def\]{\relax\ifmmode \egroup $\hfil
       \egroup \end{trivlist}\else \@badmath \fi}

\def\equation{\@beginparpenalty\predisplaypenalty
  \@endparpenalty\postdisplaypenalty
\refstepcounter{equation}\trivlist \item[]\leavevmode
  \hbox to\linewidth\bgroup $\m@th
  \displaystyle
\hskip\mathindent}

\def\endequation{$\hfil
           \displaywidth\linewidth\@eqnnum\egroup \endtrivlist}

\def\eqnarray{\stepcounter{equation}\let\@currentlabel=\theequation
\global\@eqnswtrue
\global\@eqcnt\z@\tabskip\mathindent\let\\=\@eqncr
\abovedisplayskip\topsep\ifvmode\advance\abovedisplayskip\partopsep\fi
\belowdisplayskip\abovedisplayskip
\belowdisplayshortskip\abovedisplayskip
\abovedisplayshortskip\abovedisplayskip
$$\m@th\halign
to\linewidth\bgroup\@eqnsel\hskip\@centering$\displaystyle\tabskip\z@
  {##}$&\global\@eqcnt\@ne \hskip 2\arraycolsep \hfil${##}$\hfil
  &\global\@eqcnt\tw@ \hskip 2\arraycolsep $\displaystyle{##}$\hfil
   \tabskip\@centering&\llap{##}\tabskip\z@\cr}

\def\endeqnarray{\@@eqncr\egroup
      \global\advance\c@equation\m@ne$$\global\@ignoretrue
      }

\newdimen\mathindent
\catcode`@=12

\def\lsim{\mathrel{\raise.2ex\hbox{$<$}\hskip-.8em\lower.9ex\hbox{$\sim$}}}
\def\gsim{\mathrel{\raise.2ex\hbox{$>$}\hskip-.8em\lower.9ex\hbox{$\sim$}}}

\begin{document}
\thispagestyle{empty}

\renewcommand{\thefootnote}{\fnsymbol{footnote}}

\twocolumn[
\font\fortssbx=cmssbx10 scaled \magstep1
\hbox to \hsize{
\hskip.25in \raise.05in\hbox{\fortssbx University of Wisconsin - Madison}
\hfill\vbox{\hbox{\bf MAD/PH/845}
            \hbox{August 1994}} }

\vspace{.2cm}

\begin{flushleft} {\large\bf The Case for a Kilometer-Scale High Energy
Neutrino Detector\footnotemark}\\
\vspace{.2cm}
{ F.~Halzen}\\
{\it Department of Physics, University of Wisconsin, Madison, WI 53706}\\
\end{flushleft}


{\small
Doing astronomy with photons of energies in excess of a GeV has turned out to
be extremely challenging. Efforts are underway to develop instruments that may
push astronomy to wavelengths smaller than $10^{-14}$~cm by mapping the sky in
high energy neutrinos instead. Neutrino astronomy, born with the identification
of thermonuclear fusion in the sun and the particle processes controlling the
fate of a nearby supernova, will reach outside the galaxy and make measurements
relevant to cosmology. The field is immersed in technology in the domains of
particle physics to which many of its research goals are intellectually
connected. To mind come the search for neutrino mass, cold dark matter
(supersymmetric particles?) and the monopoles of the Standard Model. While a
variety of collaborations are pioneering complementary methods by building
telescopes with effective area in excess of 0.01~km$^2$, we show here that the
natural scale of a high energy neutrino telescope is 1~km$^2$. With several
thousand optical modules and a price tag unlikely to exceed 100 million
dollars, the scope of a kilometer-scale instrument is similar to that of
experiments presently being commissioned such as the SNO neutrino observatory
in Canada and the Superkamiokande experiment in Japan.\par}

\vspace{.3cm}]

\footnotetext{Talk presented at {\it Neutrino '94, XVI International Conference
on Neutrino Physics and Astrophysics}, Eilat, Israel, May 29--June 3, 1994.}

\def\large{\normalsize}
\def\Large{\normalsize}
\renewcommand{\thesection}{\arabic{section}.}

\section*{INTRODUCTION}
\unskip
\smallskip\noindent
The photon sky has been probed with a variety of instruments sensitive to
wavelengths of light as large as $10^4$~cm for radio-waves to $10^{-14}$~cm for
the GeV-photons detected with space-based instruments. Astronomical instruments
have now collected data spanning 60 octaves in photon frequency, an amazing
expansion of the power of  our eyes which scan the sky over less than a single
octave just above $10^{-7}$~cm. Doing gamma ray astronomy at TeV energies and
beyond has, however, turned out to be a considerable challenge. Not only are
the fluxes expected to be small, as one can demonstrate by extrapolating
measured photon fluxes of MeV and GeV sources, they are dwarfed by a flux of
cosmic ray particles which is larger by typically two orders of magnitude. The
discoveries of the Crab supernova remnant and the active galaxy Markarian~421
at TeV-energy have proven that the problems are not insurmountable, more about
that later. They are, however, sufficiently daunting to have encouraged a
vigorous effort to probe the high energy sky by detecting neutrinos instead.

It is important to realize that high energy photons, unlike weakly interacting
neutrinos, do not carry information on any cosmic sites shielded from our view
by more than a few hundred grams of
intervening matter. The TeV-neutrino could reveal objects with no counterpart
in any  wavelength of light. High energy neutrinos are the decay products of
pions and are therefore a signature of the most energetic cosmic processes.
Their energy exceeds those of neutrinos artificially produced at  existing or
planned accelerators. As was the case with radiotelescopes, for instance,
unexpected discoveries could be made. Forays into new wavelength regimes have
historically led to the discovery of unanticipated phenomena. As is the case
with new accelerators, observing the predictable will be slightly
disappointing. Neutrino telescopes have the advantage of looking at a large
fraction of the sky all the time, an important astronomical advantage over
gamma ray detectors such as Cherenkov telescopes which can at best scan a few
degrees of the sky 10\% of the time. The natural background for observing
cosmic neutrinos in the TeV range and above consists of atmospheric neutrinos
which are the decay products of pions produced in cosmic ray-induced
atmospheric cascades. We will see that the expected sources actually dominate
the neutrino sky in stark contrast to the situation, previously mentioned,
where the cosmic ray flux exceeds those from high energy gamma sources by
orders of magnitude.

\section{\uppercase{Guaranteed Cosmic\hfill\break
Neutrino Beams}}

In heaven, as on Earth, high energy neutrinos are produced in beam dumps which
consist of a high energy proton (or heavy nucleus) accelerator and a target in
which gamma rays and neutrinos are generated in roughly equal numbers in the
decays of pions produced in nuclear cascades in the beam dump. For every
$\pi^0$ producing two gamma rays, there is a charged $\pi^+$ and $\pi^-$
decaying into $\mu^+\nu_\mu$. As a rule of thumb the dump produces one neutrino
for each interacting proton. If the kinematics is such that muons decay in the
dump, more neutrinos will be produced. It should be stressed immediately that
in efficient cosmic beam dumps with an abundant amount of target material, high
energy photons may be absorbed before escaping the source. Therefore, the most
spectacular neutrino sources may have no counterpart in high energy gamma rays.

By their very existence, high-energy cosmic rays do guarantee the existence of
definite sources of high energy cosmic neutrinos\cite{PR}.  They represent a
beam of known luminosity with particles accelerated to energies in excess of
$10^{20}$~eV. Cosmic rays produce pions in interactions with i)~the
interstellar gas in our galaxy, ii)~the cosmic photon background, iii)~the sun
and, finally, iv)~the Earth's atmosphere which represents a well-understood
beam dump. These interactions are the source of fluxes of diffuse photons and
neutrinos. The atmospheric neutrino beam is well understood and can be used to
study neutrino oscillations over oscillation lengths varying between 10 and
$10^4$~km\cite{PR}.

A rough estimate of the diffuse fluxes of gamma rays and neutrinos from the
galactic disk can be obtained by colliding the observed cosmic ray flux with
interstellar gas with a nominal density of 1 proton per cm$^3$. The target
material is concentrated in the disk of the galaxy and so will be the secondary
photon flux. The gamma ray flux has been identified by space-borne gamma ray
detectors. It is clear that a roughly equal diffuse neutrino flux is produced
by the decay of charged pion secondaries in the same collisions.
Conservatively, assuming a detector threshold of 1~TeV one predicts three
neutrino-induced muons per year in a $10^6\,{\rm m}^2$ detector from a solid
angle of 0.07~sr around the direction of Orion. There are several
concentrations of gas with similar or smaller density in the galaxy. The
corresponding number of neutrino events from within 10~degrees of the galactic
disc is 50~events per year for a 10$^6\,$m$^2$ detector at the South Pole which
views 1.1~steradian of the outer Galaxy with an average density of
0.013~grams/cm$^2$.

A guaranteed source of extremely energetic diffuse neutrinos is the interaction
of ultra high energy, extra-galactic, cosmic rays on the microwave background.
The major source of energy loss is photoproduction of the $\Delta$ resonance by
the cosmic proton beam on a target of background photons with a density of
$\sim$400 photons/cm$^3$ and an average energy of %
\begin{equation} \epsilon=2.7\times{\rm k_B}\times2.735^\circ \simeq
7\times10^{-4}\ {\rm eV}\ .
\end{equation}
For cosmic ray energies exceeding
\begin{equation} E_p \approx {m_{\Delta}^2 - m_p^2\over 2 (1 -\cos\theta)
\epsilon}
\approx {5 \times 10^{20} \over (1 - \cos\theta)}\; {\rm eV} \; ,
\end{equation}
where $\theta$ is the angle between the proton and photon directions, the
photopion cross-section grows very rapidly to reach a maximum of 540~$\mu$b at
the $\Delta^+$ resonance $(s = 1.52$~GeV$^2)$. The $\Delta^+$ decays to $p
\pi^0$ with probability of 2/3, and to $n\pi^+$ with probability 1/3. The
charged pions are the source of very high energy muon-neutrino fluxes as a
result of decay kinematics where each neutrino takes approximately 1/4 of the
parent pion energy. In addition neutrons decay producing a flux of lower energy
$\bar\nu_e$.

The magnitude and intensity of the cosmological neutrino fluxes is determined
by the maximum injection energy of the ultra-high-energy cosmic rays and by the
distribution of their sources. If the sources are relatively near at distances
of order tens of Mpc, and the maximum injection energy is not much greater than
the highest observed cosmic ray energy (few $\times 10^{20}$~eV), the generated
neutrino fluxes are small. If, however, the highest energy cosmic rays are
generated by many sources at large redshift, then a large fraction of their
injection energy would be presently contained in $\gamma$-ray and neutrino
fluxes. The reason is that the energy density of the microwave radiation as
well as the photoproduction cross-section scale as $(1 + z)^4$. The effect
would be even stronger if the source luminosity were increasing with $z$, i.e.\
cosmic ray sources were more active at large redshifts -- `bright phase'
models. Early speculations on bright phase models led to the suggestion of
kilometer-scale neutrino detectors over a decade ago\cite{Venya}.

The other guaranteed extraterrestrial source of high energy neutrinos is the
Sun. The production process is exactly the same as for atmospheric neutrinos on
Earth: cosmic ray interactions in the solar atmosphere. Neutrino production is
enhanced because the atmosphere of the Sun is much
more tenuous. The scaleheight of the chromosphere is $\sim$115~km, compared
with 6.3~km for our upper atmosphere. A detailed calculation of the neutrino
production by cosmic rays in the solar atmosphere shows a neutrino spectrum
larger than the angle averaged atmospheric flux by a factor of $\sim$2 at
10~GeV and a factor of $\sim$3 at 1000~GeV. The decisive factor for the
observability of this neutrino source is the small solid angle $(6.8\times
10^{-5}$ sr) of the Sun. Although the rate of the neutrino induced upward going
muons is higher than the atmospheric emission from the same solid angle by a
factor of $\sim$5, the rate of muons of energy above 10~GeV in a $10^6$ m$^2$
detector is only 50 per year. Taking into account the diffusion of the cosmic
rays in the solar wind, which decreases the value of the flux for energies
below one TeV, cuts this event rate by a factor of 3. Folded with a realistic
angular resolution of 1 degree, observation of such an event rate requires, as
for the previous examples, a 1~km$^2$ detector.

\medskip\smallskip

\section{\uppercase{Active Galactic Nuclei:\hfil\break
Almost Guaranteed?}}

Although observations of PeV ($10^{15}$~eV) and EeV ($10^{18}$~eV) gamma-rays
are controversial, cosmic rays of such energies do exist and their origin is at
present a mystery. The cosmic-ray spectrum can be understood, up to perhaps
1000~TeV, in terms of shockwave acceleration in galactic supernova remnants.
Although the spectrum suddenly steepens at
1000~TeV, a break usually referred  to as the ``knee\rlap,'' cosmic rays with
much higher energies are observed and cannot be accounted for by this
mechanism. This failure can be understood by simple dimensional analysis
because the EMF in the supernova shock is of the form
\begin{equation}
E = ZeBRc \,,
\label{SNEMF}
\end{equation}
where $B$ and $R$ are the magnetic field and the radius of the shock. For a
proton Eq.~(\ref{SNEMF}) yields a maximum energy
\begin{equation} E_{\rm max} = \left[10^5\,{\rm
TeV}\right]\left[B\over3\times10^{-6}\rm\,G\right]
\left[R\over50\,{\rm pc}\right]
\end{equation}
and therefore $E$ is less than $10^5$~TeV for the typical values of $B,R$
shown. The actual upper limit is much smaller than the value obtained by
dimensional analysis because of inefficiencies in the acceleration process.

Cosmic rays with energy in excess of $10^{20}$~eV have been observed. Assuming
that they are a galactic phenomenon, the measured spectrum implies that
$10^{34}$ particles are accelerated to
1000~TeV energy every second. We do not know where or how. We do not know
whether the particles are protons or iron or something else. If the cosmic
accelerators indeed  exploit the $3 \mu$Gauss field of our galaxy, they must be
much larger than supernova remnants in order to reach $10^{21}$~eV energies.
Equation~(\ref{SNEMF}) requires that their size be of order 30~kpc. Such an
accelerator exceeds the dimensions of our galaxy. Although imaginative
arguments exist to avoid this impasse, an attractive alternative is to look for
large size accelerators outside the galaxy. Nearby active galactic nuclei
(quasars, blazars\dots) distant by order 100~Mpc are the obvious candidates.
With magnetic fields of tens of $\mu$Gauss over distances of kpc near the
central black hole or in the jets, acceleration to $10^{21}$~eV is possible;
see Eq.~(\ref{SNEMF}).

One can visualize the accelerator in a very economical way in the
Blanford-Zralek mechanism. Imagine that the horizon of the central black hole
acts as a rotating conductor immersed in an external magnetic field. By simple
dimensional analysis this creates a voltage drop
\begin{equation} {\Delta V\over 10^{20}{\rm volts}} = {a\over M_{\rm BH}} \,
{B\over 10^4{\rm G}} \, {M_{\rm BH}\over 10^9 M_\odot} \;,
\end{equation}
corresponding to a luminosity
\begin{equation} {{\cal L}\over10^{45}\rm erg\,s^{-1}} \! = \!\! \left(a\over
M_{\rm BH}\right)^2 \!\!
\left(B\over 10^4\rm G\right)^2\!\! \left(M_{\rm BH}\over
10^9M_\odot\right)^2\!\!.
\end{equation}
Here $a$ is the angular momentum per unit mass of a black hole of mass $M_{\rm
BH}$.

All this was pretty much a theorist's pipe dream until recently the Whipple
collaboration reported the observation of TeV ($10^{12}\,$eV) photons from the
giant elliptical galaxy Markarian~421\cite{Mkr}. With a signal in excess of
6~standard deviations, this was the first convincing observation of TeV gamma
rays from outside our Galaxy. That a distant source such as Markarian~421 can
be observed at all implies that its luminosity exceeds that of galactic cosmic
accelerators such as the Crab, the only source observed by the same instrument
with comparable statistical significance, by close to 10 orders of magnitude.
More distant by a factor $10^5$, the instrument's solid angle for Markarian~421
is reduced by $10^{-10}$ compared to the Crab. Nevertheless the photon count at
TeV energy is roughly the same for the two sources. The
Whipple observation implies a Markarian~421 photon luminosity in excess of
$10^{43}$ ergs per second. It is interesting that these sources have their
highest luminosity above TeV energy, beyond the wavelengths of conventional
astronomy. During May 1994 observations of the flux of Markarian~421 was
observed to increase by a factor 10 in one day, strongly suggesting the
catastrophic operation of a high energy hadronic accelerator.

\medskip\smallskip

Why Markarian 421? Whipple obviously zoomed in on the Compton Observatory
catalogue of active galaxies (AGNs) known to emit GeV photons. Markarian, at a
distance of barely over 100~Mpc, is the closest blazar on the list. As yet TeV
gamma rays have not been detected from any other AGNs. Although Markarian 421
is the closest of these AGNs, it is one of the weakest; the reason that it is
detected whereas other, more distant, but more powerful, AGNs are not, must be
that the TeV gamma rays suffer absorption in intergalactic space through the
interaction with background infra-red photons. TeV gamma rays are indeed
efficiently absorbed on infra-red starlight and this most likely provides the
explanation why astronomers have a hard time observing much more powerful
quasars such as 3C279 at a redshift of 0.54. Production of $e^+e^-$ pairs by
TeV gamma rays interacting with IR background photons is the origin of the
absorption. The absorption is, however, minimal for Mrk~421 with $z=0.03$, a
distance close enough to see through the IR~fog. This implies that all of the
AGNs may have significant very high energy components but that only Markarian
421 is close enough to be\break
 detectable with currently available gamma-ray telescopes.

This observation was not totally unanticipated. Many theorists\cite{PR} have
identified blazars such as Markarian~421 as powerful cosmic accelerators
producing beams of very high energy photons and neutrinos. Acceleration of
particles is by shocks in the jets (or, possibly, also by shocks in the
accretion flow onto the supermassive black hole which powers the galaxy) which
are a characteristic feature of these radio-loud active galaxies. Many
arguments have been given for the acceleration of protons as well as electrons.
Inevitably beams of gamma rays and neutrinos from the decay of pions appear
along the jets. The pions are photoproduced by accelerated protons on the
target of optical and UV~photons in the galaxy which reaches densities of
$10^{14}$ per cm$^3$. The latter are the product of synchrotron radiation by
electrons accelerated along with the protons.

Powerful AGNs at distances of order 100 Mpc and with proton luminosities of
$10^{45}\,$erg/s or higher are obvious candidates for the cosmic accelerators
of the highest energy cosmic rays. Their luminosity often peaks at the highest
energies and their proton flux, propagated to Earth, can quantitatively
reproduce the cosmic ray spectrum above spectrum $10^{18}$~eV\cite{bier}. Some
have argued that all cosmic rays above the ``knee" in the spectrum at
$10^{15}$~eV may be of AGN origin. The neutrino flux from such accelerators can
be calculated by energy
conservation:
\begin{equation} {\cal L}_p N\epsilon_{\rm eff} = 4\pi d^2 \int dE [E\, dN_\nu
/ dE] \;,
\end{equation}
where $N_\nu$ is the neutrino flux at Earth, $d$ the average distance to the
sources, $N$ the number of sources and $\epsilon_{\rm eff}$ the efficiency for
protons to produce pions (or neutrinos, assuming the production of 1~neutrino
per interacting proton) in the AGN beamdump. This yields
\begin{equation} E {dN_\nu\over dE} = {N \epsilon_{\rm eff}\over 4\pi}
{7.5\times 10^{-10} \over E\,\rm(TeV)} \,\rm cm^{-2}\, s^{-1}\, sr^{-1}
\label{flux}
\end{equation}
for ${\cal L}_p=10^{45}\,$erg/s and $d=100$~Mpc. We here assumed an $E^{-2}$
energy spectrum extending to $10^{20}$~eV energy. With $\epsilon_{\rm eff}$ of
order $10^{-1}$ to $10^{-3}$ and the number of relatively nearby sources $N$ in
the range 10 to 1000, it is a reasonable estimate that $N\epsilon_{\rm eff}=1$.
The total energy in excess of 1~EeV ($10^{18}$-$10^{20}$~eV) is $5\times
10^{-9}$ erg/cm$^2$/s. This number nicely matches the energy density of the
extra-galactic cosmic rays in the same interval of energy as it should
assuming, again, that 1~neutrino is produced for every proton in the AGN dump.
The flux of Eq.~(\ref{flux}) is at the low end of the range of fluxes predicted
by Biermann et al. and by Protheroe et al.\ and Stecker et al.\ in models where
acceleration is in shocks in the jet\cite{bier} and accretion
disc\cite{Ray,Floyd}, respectively.

The above discussion suggests a very simple estimate of the AGN neutrino flux
that finesses all guesses regarding the properties of individual sources:
\begin{equation}
\begin{array}{l} \displaystyle
4\pi \int_{10^{17}\,\rm eV} dE [ E\,dN_\nu/dE] \simeq {\cal L}_{\rm CR} \\
\\
\quad\simeq 7.2\times10^{-9}\,\rm erg\,cm^{-2}\,s^{-1} \;,
\end{array}
\end{equation}
which simply states that AGNs generate 1 neutrino for each proton. ${\cal
L}_{\rm CR}$ is obtained by integrating the highest energy $E^{-2.71}$
component of the cosmic ray flux above $10^{17}$\,eV. Assuming an $E^{-2}$
neutrino spectrum we recover the result of Eq.~(\ref{flux}). Is is now clear
that our flux is a lower limit as protons should be absorbed in ambient matter
in the source or in the interstellar medium.

\section{\uppercase{Intermezzo: The Case for
 a Kilometer-Scale Detector}}

Observing AGNs has become a pivotal goal in the development of high energy
neutrino telescopes. Neutrinos are observed via the muons they produce in the
detector volume. At high energy it is possible to enhance the effective volume
of detectors by looking for neutrino-induced muons generated in charged-current
interactions of $\nu_\mu$ in the water or ice outside the instrumented detector
volume. The effective detector volume is then the product of the detector area
and the muon range in rock R$_\mu$. TeV muons have a typical range of one
kilometer, which leads to a significant increase in effective detector volume.
The average muon energy loss rate is
\begin{equation} {\left\langle{\rm d}E\over{\rm d}X\right\rangle}\;=\;
-\alpha(E)\,-\,\beta(E)\times E\;,
\end{equation}
where $X$ is the thickness of material in g/cm$^2$. The first term represents
ionization losses, which are approximately independent of energy, with
$\alpha\sim 2$~MeV~g$^{-1}$cm$^2$. The second term includes the catastrophic
processes of bremsstrahlung, pair production and nuclear interactions, for
which fluctuations play an essential role.  Here $\beta\sim 4\times
10^{-6}$~g$^{-1}$cm$^2$. The critical energy above which the radiative
processes
dominate is
\begin{equation} E_{\rm cr}\;=\;\alpha/\beta\;\approx\;500\;{\rm GeV}.
\end{equation}
To treat muon propagation properly when $E_\mu>E_{\rm cr}$ requires a Monte
Carlo calculation of the probability $P_{\rm surv}$ that a muon of energy
$E_\mu$ survives with energy $>E_\mu^{\rm min}$ after propagating a distance
$X$.  The probability that a neutrino of energy $E_\nu$ on a trajectory through
a detector produces a muon above threshold at the detector is
\begin{equation}
\begin{array}{rl}
 P_\nu(E_\nu,E_\mu^{\rm min})
\displaystyle  = N_A\,\int_0^{E_\nu}\!\!\!\!\!\! & \!\!\!\!\!\!{\rm d}
E_\mu{{\rm d}\sigma_\nu\over{\rm d}E_\mu}(E_\mu,E_\nu)\\
&\times R_{\rm eff}(E_\mu,E_\mu^{\rm min})\;,
\end{array}
\label{P nu E}
\end{equation}
where
\begin{equation} R_{\rm eff}\,=\,\int_0^\infty
\,{\rm d}X\,P_{\rm surv}(E_\mu,E_\mu^{\rm min},X)\;.
\end{equation}
The flux of $\nu_\mu$-induced muons at the detector is given by a convolution
of the neutrino spectrum $\phi_\nu$ with the muon production probability
(\ref{P nu E}) as
\begin{equation}
\begin{array}{l} \displaystyle
\phi_\mu(E_\mu^{\rm min},\theta)
 = \int_{E_\mu^{\rm min}}  \Bigl\{  {\rm d}
E_\nu\,P_\nu(E_\nu,E_\mu^{\rm min})\\ \noalign{\vskip1ex}
\hspace*{2.5em} \times \exp[-\sigma_{\rm t}(E_\nu)\,X(\theta)\,N_A]\,
\phi_\nu(E_\nu,\theta) \Bigr\} \;.
\end{array}
\label{$N_mu$}
\end{equation}
The exponential factor here accounts for absorption of neutrinos along the
chord of the Earth, $X(\theta)$.  Absorption becomes important for
$\sigma(E_\nu)\gsim 10^{-33}$~cm$^2$
or $E_\nu\gsim 10^7$~GeV.

The event rate in a detector is obtained by multiplying Eq.~(\ref{$N_mu$}) by
its effective area. From Eqs.~(\ref{flux}),(\ref{$N_mu$}) we obtain order 300
upcoming muon events per year in a $10^6\,{\rm m}^2$ detector. It is not a
comfortably large rate however as the flux is indeed distributed over a large
number of sources. There is, however, no competing background. Hopefully one
will be able to scrutinize a few nearby sources with good statistics. We should
recall at this point that our back-of-the-envelope estimate yields a flux at
the lower end of the range of fluxes predicted by detailed modeling. Optimistic
predictions exceed our estimate by over an order of magnitude and are,
possibly, within reach of the telescopes now being commissioned.

The neutrino sky at GeV-energy and above is summarized in Fig.~1. Shown is the
flux from the galactic plane as well as a range of estimates (from generous to
conservative) for the diffuse fluxes of neutrinos from active galaxies and from
the interaction of extra-galactic cosmic rays with cosmic photons. At PeV
energies and above all sources dominate the background of atmospheric
neutrinos. In order to deduce the effective area of an instrument required to
study the fluxes in the figure, the detection efficiency must be included using
Eq.~(\ref{$N_mu$}). At the highest energies this efficiency approaches unity
and 1 event per km$^2$ per year corresponds to the naive estimate of $10^{-18}$
neutrinos per cm$^2$ second. At TeV--PeV energy the one event level per year
corresponds to a flux of $10^{-14}$--$10^{-15}$ per cm$^2$ second. As before,
we conclude that the diffuse flux from AGN yields order $10^3$ events in a
kilometer-size detector per year in the TeV-energy range.

\begin{figure}[h]
\centering
\epsfxsize=7.5cm
\hspace{0in}\epsffile{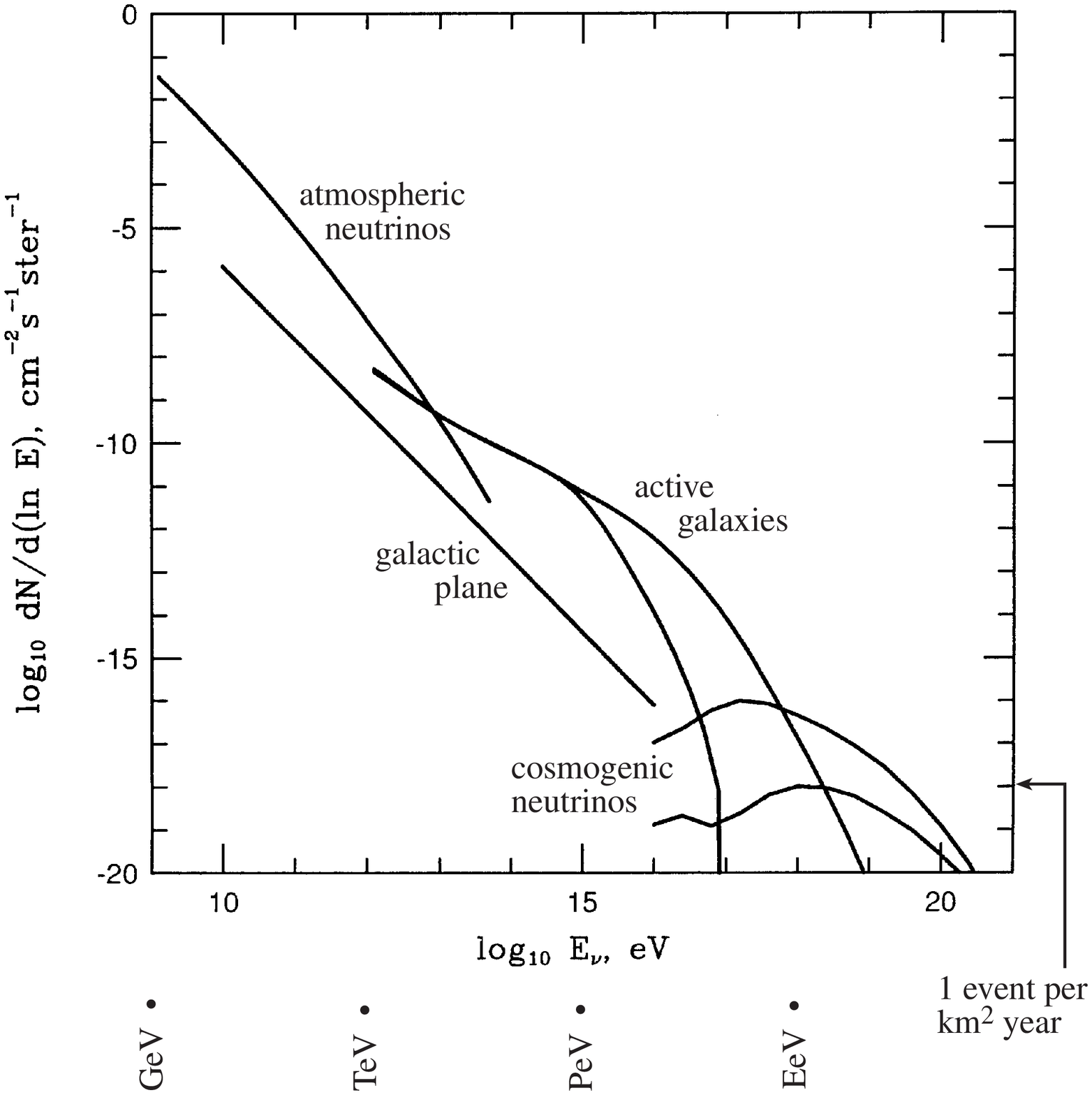}

\medskip
\small Fig.~1
\end{figure}

It should be emphasized that high energy neutrino detectors are multi-purpose
instruments. Their science-reach touches astronomy, astrophysics and particle
physics. Further motivations for the construction of a km$^3$ deep underground
detector include\cite{PR}:

\begin{enumerate}
\addtolength{\itemsep}{-.1in}
\item
The search for the t'Hooft-Polyakov mono\-poles predicted by the Standard
Model.

\item
The study of neutrino oscillations by monitoring the atmospheric neutrino beam.
One can exploit the unique capability of relatively shallow neutrino
telescopes, i.e.\ detectors positioned at a depth of roughly 1~km, to detect
neutrinos and muons of similar energy. In a $\nu_{\mu}$ oscillation experiment
one can therefore tag the $\pi$ progenitor of the neutrino by detecting the
muon produced in the same decay. This eliminates the model dependence of the
measurement inevitably associated with the calculation of the primary cosmic
ray flux. Surface neutrino telescopes probe the parameter space $\Delta m^2
\gsim 10^{-3}\rm\ eV^2$ and $\sin^2 2\theta \gsim 10^{-3}$ using this
technique. Recently underground experiments have given tantalizing hints for
neutrino oscillations in this mass range.

\item
The search for neutrinos from the annihilation of dark matter particles in our
galaxy.

\item
The capability to observe the thermal neutrino emission from
supernovae\cite{Super} (even though the nominal threshold of the detectors
exceeds the neutrino energy by several orders of magnitude!). The detector will
be able to monitor our galaxy over decades in a most economical fashion.

\item
Further study of the science pioneered by space-based gamma ray detectors such
as the study of gamma ray bursts and the high energy emission from quasars.

\end{enumerate}

It is intriguing that each of these goals point individually at the necessity
to commission a kilometer-size detector. In order to illustrate this I will
discuss the search for the particles which constitute the cold dark matter.

\section{\uppercase{Indirect Search for\hfill\break
 Cold Dark Matter}}

It is believed that most of our Universe is made of cold dark matter particles.
Big bang cosmology implies that these particles have interactions of order the
weak scale, i.e.\ they are WIMPs\cite{SeckelDM}. We know everything about these
particles except whether they really exist. We know that their mass is of order
of the weak boson mass, we know that they interact weakly. We also know their
density and average velocity in our galaxy as they must constitute the dominant
component of the density of our galactic halo as measured by rotation curves.
WIMPs will annihilate into neutrinos with rates whose estimate is
straightforward. Massive WIMPs will annihilate into high energy neutrinos.
Their detection by high energy neutrino telescopes is greatly facilitated by
the fact that the sun, conveniently, represents a dense and nearby source of
cold dark matter particles.

Galactic WIMPs, scattering off protons in the sun, lose energy. They may fall
below escape velocity and be gravitationally trapped. Trapped dark matter
particles eventually come to equilibrium temperature, and therefore to rest at
the center of the sun. While the WIMP density
builds up, their annihilation rate into lighter particles increases until
equilibrium is achieved where the annihilation rate equals half of the capture
rate. The sun has thus become a reservoir of WIMPs which annihilate into any
open fermion, gauge boson or Higgs channels. The leptonic decays from
annihilation channels such as $b\bar b$ heavy quark and $W^+W^-$ pairs turn the
sun into a source of high energy neutrinos. Their energies are in the GeV to
TeV range, rather than in the familiar KeV to MeV range from its thermonuclear
burning. These neutrinos can be detected in deep underground experiments.

We illustrate the power of neutrino telescopes as dark matter detectors using
as an example the search for a 500~GeV WIMP with a mass outside  the reach of
present accelerator and future LHC experiments. A quantitative estimate of the
rate of high energy muons of WIMP origin triggering a detector can be made in 5
easy steps.

\smallskip
\leftline{{\bf Step 1:} The halo neutralino flux $\phi_{\chi}$.}

\noindent
It is given by their number density  and average velocity. The cold dark matter
density implied by the observed galactic rotation curves is $\rho_\chi = 0.4$
GeV/cm$^3$. The galactic halo is believed to be an isothermal sphere of WIMPs
with average velocity $v_\chi = 300$~km/sec. The number density is then
\begin{equation} n_\chi = 8\times 10^{-4} \left[ 500{\rm\ GeV}\over m_\chi
\right]\rm\
cm^{-3}
\end{equation}
and therefore
\begin{equation} \phi_\chi = n_\chi v_\chi = 2\times 10^{4} \left[ 500{\rm\
GeV}\over
m_\chi \right] \rm\ cm^{-2}\, s^{-1} \,. \label{fluxprime}
\end{equation}

\noindent
{\bf Step 2:} Cross section $\sigma_{\rm sun}$ for the capture of neutralinos
by the sun.

\noindent
The probability that a WIMP is captured is proportional to the number of target
hydrogen nuclei in the sun (i.e.\ the solar mass divided by the nucleon mass)
and the WIMP-nucleon scattering cross section. From dimensional analysis
$\sigma(\chi N)\; \sim \; \left(G_F m_N^2\right)^2/m_Z^2$ which we can envisage
as the exchange of a neutral weak boson between the WIMP and a quark in the
nucleon. The main point is that the WIMP is known to be weakly interacting. We
obtain for the solar capture cross section
\begin{equation}
\begin{array}{rl} \displaystyle
 \Sigma_{\rm sun} = n\sigma \!\!\!\!&= {M_{\rm sun}\over m_N} \sigma(\chi N)\\
\noalign{\vskip1ex}
&= \left[1.2\times 10^{57}\right] \left[10^{-41}\,\rm cm^2\right] \,.
\end{array}
\label{capture}
\end{equation}

\noindent
{\bf Step 3:} Capture rate $N_{\rm cap}$ of neutralinos by the sun.

\noindent
$N_{\rm cap}$ is determined by the WIMP flux (\ref{fluxprime}) and the sun's
capture cross section (\ref{capture}) obtained in the first 2 steps:
\begin{equation} N_{\rm cap} = \phi_\chi \Sigma_{\rm sun} = 3\times
10^{20}\,\rm
s^{-1} \,.
\end{equation}

\noindent
{\bf Step 4:} Number of solar neutrinos of dark matter origin.

\noindent
The sun comes to a steady state where capture and annihilation of WIMPs are in
equilibrium. For a 500~GeV WIMP the dominant annihilation rate is into weak
bosons; each produces muon-neutrinos with a branching ratio which is roughly
10\%:
\begin{equation}\chi\bar\chi \to WW \to \mu \nu_\mu \,. \label{branch}
\end{equation}
Therefore, as we get 2 $W$'s for each capture, the number of  neutrinos
generated in the sun is
\begin{equation} N_\nu = {1\over 5} N_{\rm cap}
\end{equation}
and the corresponding neutrino flux at Earth is given by
\begin{equation}  \phi_\nu = {N_\nu\over 4\pi d^2} = 2\times 10^{-8}\,\rm
cm^{-2}
s^{-1} \,,
\label{nu flux}
\end{equation}
where the distance $d$ is 1 astronomical unit.

\goodbreak
\noindent
{\bf Step 5:} Event rate in a high energy neutrino telescope.

\noindent
For (\ref{branch}) the $W$-energy is approximately $m_{\chi}$ and the neutrino
energy half that by two-body kinematics. The energy of the detected muon is
given by
\begin{equation}  E_\mu \simeq {1\over2} E_\nu \simeq {1\over4}m_\chi \,.
\end{equation}
Here we used the fact that, in this energy range, roughly half of the neutrino
energy is transferred to the muon. Simple estimates of the neutrino
interaction cross section and the muon range can be obtained as follows
\begin{equation}
\sigma_{\nu\to\mu} = 10^{-38}\,{\rm cm^2}\,{E_\nu\over \rm GeV} =
2.5\times10^{-36}\,\rm cm^2
\end{equation}
and
\begin{equation}
R_\mu = 5\ {\rm m}\,{E_\mu\over\rm GeV} = 625\ \rm m \,,
\end{equation}
which is the distance covered by a muon given that it loses 2~MeV for each
gram of matter traversed. We have now collected all the information to compute
the number of events in a detector of area $10^4\,$m$^2$, typical for those
presently under construction.

For the neutrino flux given by (\ref{nu flux}) we obtain
\begin{eqnarray}
 {\rm \#\,events/year } \!\!\!\!&=&\!\!\!\! 10^6 \times \phi_\nu \times
\rho_{\rm H_2O}
\times \sigma_{\nu\to\mu} \times R_\mu \nonumber\\
\!\!\!\!&\simeq&\!\!\!\! 1000
\end{eqnarray}
for a 1~km$^2$ water Cherenkov detector, where $R_\mu$ is the
muon range and $\phi_\nu \times \rho_{\rm H_2O} \times
\sigma_{\nu\to\mu}$ is the analog of Eq.~(\ref{$N_mu$}).

The above exercise is just meant to illustrate that high energy neutrino
telescopes compete with present and future accelerator experiments in  the
search for dark matter and
supersymmetry; see below. The above exercise can be repeated as a function of
WIMP mass. The result is shown in Fig.~2 (the two branches as well as the
structure in the curves are related to details of supersymmetry. These are, for
all practical purposes, irrelevant). Especially for heavier WIMPs the technique
is very powerful because underground high energy neutrino detectors have been
optimized to be sensitive in the energy region where the neutrino interaction
cross section and the range of the muon are large.  Also, for high energy
neutrinos the muon and neutrino are aligned along a direction  pointing back to
the sun with good angular resolution. A kilometer-size detector probes WIMP
masses up to the TeV-range beyond which they are excluded by cosmological
considerations. The technique fails for low masses only for those mass values
already excluded by unsuccessful accelerator searches. Competitive direct
searches for dark matter will have to deliver detectors reaching better than
0.05~events/kg/day \hbox{sensitivity.}

\begin{figure}

\epsfxsize=7.5cm
\hspace{0in}\epsffile{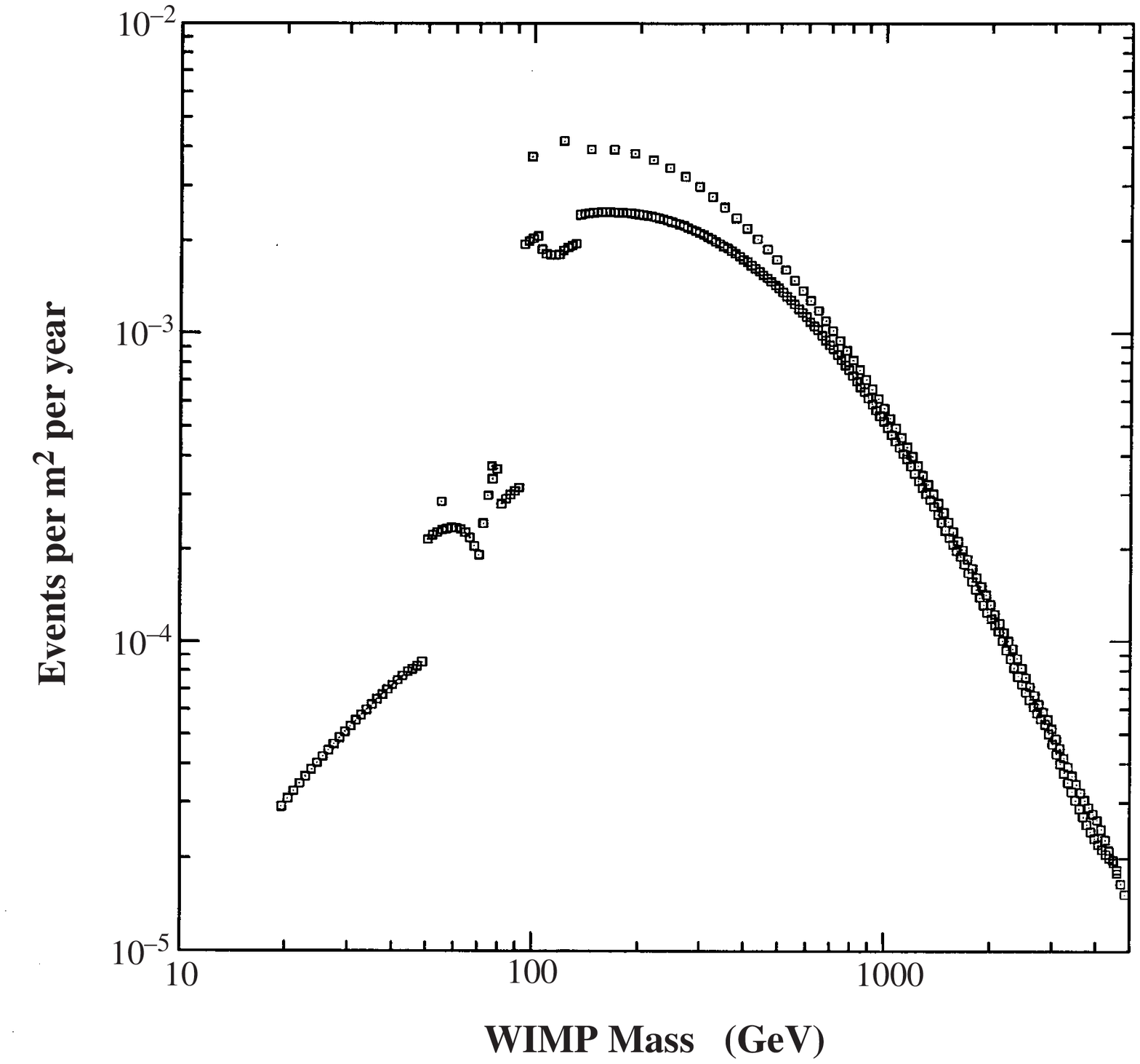}\hspace*{.25in}

\medskip
\small\centering Fig.~2
\vspace*{-.5cm}

\end{figure}

Particle physics provides us with rather compelling candidates for WIMPs. The
Standard Model is not a model: its radiative corrections are not under control.
A most elegant and economical way to revamp it into a consistent and calculable
framework is to make the model supersymmetric. If supersymmetry is indeed
Nature's extension of the Standard Model it must produce new phenomena at or
below the TeV scale. A very attractive feature of supersymmetry is that it
provides cosmology with a natural dark matter candidate in form of a stable
lightest supersymmetric particle\cite{SeckelDM}. This is, in fact, the only
candidate because supersymmetry completes the Standard Model all the way to the
GUT scale where its forces apparently unify. Because supersymmetry logically
completes the Standard Model with no other new physics threshold up to the
GUT-scale, it must supply the dark matter. So, if supersymmetry, dark matter
and accelerator detectors are on a level playing field. The interpretation of
above arguments in the framework of supersymmetry are explicitly stated in
Ref.~\cite{Kamionkowski}.

\section{\uppercase{DUMAND et al.: Complementary Technologies}}

We have presented arguments for doing neutrino astronomy on the scale of
1~kilometer. In order to achieve large effective area it is, unfortunately,
necessary to abandon the low MeV thresholds of detectors such as IMB and
Kamiokande. One focuses on high energies where: i) neutrino cross sections are
large and the muon range is increased; see Equation~(\ref{$N_mu$}), ii) the
angle between the muon and parent neutrino is less than 1~degree and, iii) the
atmospheric neutrino background is small.
The accelerator physicist's method for building a neutrino detector uses
absorber, chambers with a few $x,y$ wires and associated electronics with a
price of roughly $10^4$~US dollars per m$^2$. Such a 1~km$^2$ detector would
cost 10~billion dollars.  Realistically, we are compelled to develop methods
which are more cost-effective by a factor 100 in order to be able to commission
neutrino telescopes with area of order 1~km$^2$. Obviously, the proven
technique developed by IMB, Kamiokande and others cannot be extrapolated to
kilometer scale. All present telescopes do however exploit the well-proven
Cherenkov technique.

In a Cherenkov detector the direction of the neutrino is inferred from the muon
track which is measured by mapping the associated Cherenkov cone travelling
through the detector. The arrival times and amplitudes of the Cherenkov
photons, recorded by a grid of optical detectors, are used to reconstruct the
direction of the radiating muon. The challenge is well-defined: record the muon
direction with sufficient precision (i.e., sufficient to reject the much more
numerous down-going cosmic ray muons from the up-coming muons of neutrino
origin) with a minimum number of optical modules (OM). Critical parameters are
detector depth which determines the level of the cosmic ray muon background and
the noise rates in the optical modules which will sprinkle a muon trigger with
false signals. Sources of such noise include radioactive decays such as
potassium decay in water, bioluminescence and, inevitably, the dark current of
the photomultiplier tube. The experimental advantages and challenges are
different for each experiment and, in this sense, they nicely complement one
another as engineering projects for a large detector. Each has its own
``gimmick'' to achieve neutrino detection with a minimum number of OMs:

\begin{enumerate}
\addtolength{\itemsep}{-0.05in}

\item AMANDA uses sterile ice, free of radioactivity;

\item Baikal triggers on pairs of OMs;

\item DUMAND and NESTOR shield their arrays by over 4~km of ocean water.

\end{enumerate}

Detectors under construction will have a nominal effective area of
$10^4$~m$^2$. The OMs are deployed like beads on strings separated by
20--50~meters.  There are typically 20 OMs per string separated by roughly
10~meters. Baikal is presently operating 36 optical modules, 18 pointing up and
18 down, and the South Pole AMANDA experiment started operating 4 strings with
20 optical modules each in January 94. The first generation telescopes will
consist of roughly 200 OMs. Briefly,

\begin{enumerate}
\addtolength{\itemsep}{-0.05in}
\item
AMANDA is operating in deep clear ice with an absorption length in excess of
60~m similar to that of the clearest water used in the Kamiokande and IMB
experiments. The ice provides a convenient mechanical support for the detector.
The immediate advantage is that all electronics can be positioned at the
surface. Only the  optical modules are deployed into the deep ice. Polar ice is
a sterile medium with a concentration of radioactive elements reduced by more
than $10^{-4}$ compared to sea or lake water. The low background results in an
improved sensitivity which allows for the detection of high energy muons with
very simple trigger schemes which are implemented by off-the-shelf electronics.
Being  positioned under only 1~km of ice it is operating in a
cosmic ray muon  background which is over 100 times larger than deep-ocean
detectors such as DUMAND. The challenge is to reject the down-going muon
background relative to the up-coming neutrino-induced muons by a factor larger
than $10^6$. The group claims to have met this challenge with an up/down
rejection which is similar to that of the deep detectors.

Although residual bubbles are found at depth as large as 1~km, their density
decreases rapidly with depth.  Ice at the South Pole should be bubble-free
below 1100--1300~m as it is in other polar regions. The effect of bubbles on
timing of photons has been measured by the laser calibration system deployed
along with the OMs. After taking the scattering of the light on bubbles into
account reconstruction of muons has been demonstrated by a successful
measurement of the characteristic fluxes of cosmic ray muons.

The polar environment turned out to be surprisingly friendly but only allows
for restricted access and one-shot deployment of photomultiplier strings. The
technology has, however, been satisfactorily demonstrated with the deployment
of the first 4 strings. It is clear that the hot water drilling  technique can
be used to deploy OM's larger than the 8~inch photomultiplier tubes now used to
any depth in the 3~km deep ice cover. AMANDA will deploy 6 more strings in 1995
at a depth of 1500~meters.

\item
BAIKAL shares the shallow depth of AMANDA and large background counting rate of
tens of kHz from bioluminescence and radioactive decays with DUMAND. It
suppresses its background by pairing OMs in the trigger. Half its optical
modules are pointing up in order to achieve a uniform acceptance over upper and
lower hemispheres. The depth of the lake is 1.4~km, so the experiment cannot
expand downwards and will have to grow  horizontally.

The Baikal group has been operating an array of 18(36) Quasar photomultiplier
(a Russian-made 15~inch tube) units deployed in April 1993(94). They have
reached a record up/down rejection ratio of $10^{-4}$ and, according to Monte
Carlo, will reach the $10^{-6}$ goal to detect neutrinos as soon as the full
complement of 200 OMs is deployed. They expect to deploy 97 additional OMs in
1995.

\item
DUMAND will be positioned under 4.5~km of ocean water, below most biological
activity and well shielded from cosmic ray muon backgrounds. A handicap of
using ocean water is the background light resulting from radioactive decays,
mostly K$^{40}$, plus some bioluminescence, yielding a noise rate of 60~kHz in
a single OM. Deep ocean  water is, on the other hand very clear, with an
attenuation length of order 40~m in the blue. The deep ocean is a difficult
location for access and service. Detection equipment must be built to high
reliability standards, and the data must be transmitted to the shore station
for processing.  It has required years to develop the necessary technology and
learn to work in an environment foreign to high-energy physics experimentation,
but  hopefully this will be accomplished satisfactorily.

The DUMAND group has successfully analysed data on cosmic ray muons from the
deployment of a test string. They have already installed the 25~km power and
signal cables from detector to shore as well as the junction box for deploying
the strings. The group will proceed with the
deployment of 3 strings in 1995.

\item
NESTOR is similar to DUMAND, being placed in the deep ocean (the
Mediterranean), except for two critical differences. Half of its optical
modules point up, half down like Baikal. The angular
response of the detector is  being tuned to be much more isotropic than either
AMANDA or DUMAND, which  will give it advantages in, for instance, the study of
neutrino oscillations.  Secondly, NESTOR will have a higher density of
photocathode (in some substantial volume) than the other detectors, and will be
able to make local coincidences on lower energy events, even perhaps down to
the supernova energy range (tens of MeV).

\item
Other detectors have been proposed for near surface lakes or ponds (e.g.
GRANDE, LENA, NET, PAN and the Blue Lake Project), but at this time none are in
construction\cite{Learnedprime}.  These detectors all  would have the great
advantage of accessibility and ability for dual use as  extensive air shower
detectors, but suffer from the $10^{10}$--10$^{11}$ down-to-up ratio of muons,
and face great civil engineering costs (for water systems and light-tight
containers). Even if any of these are built it would seem  that the costs may
be too large to contemplate a full kilometer-scale detector.

\end{enumerate}

\section{\uppercase{Sketch of a Kilometer-Size Detector}}

In summary, there are four major experiments proceeding with construction, each
of which has different strengths and faces different challenges.  For the
construction of a 1~km scale detector one can imagine any of the above
detectors being the basic building block for the ultimate 1~km$^3$ telescope.
The redesigned AMANDA detector (with spacings optimized to the absorption
length of 60~m), for example, consists of 5 strings on a circle of 60~meter
radius around a string at the center (referred to as a $1+5$ configuration).
Each string contains 13 OMs separated by 15~m. Its effective volume for
TeV-neutrinos is just below $10^7$~m$^3$. Imagine AMANDA ``supermodules'' which
are obtained by extending the basic string length (and module count per string)
by a factor close to 4. Supermodules would then consist of $1+5$ strings with
51 OMs
separated by 20~meters on each string, for a total length of 1~km. A 1~km scale
detector then may consist of a $1+7+7$ configuration of supermodules, with the
7 supermodules distributed on a circle of radius 250~m and 7 more on a circle
of 500~m. The full detector then contains 4590 phototubes, which is less than
the 9000 used in the SNO detector. Such a detector (see Fig.~3) can be operated
in a dual mode:

\begin{enumerate}
\addtolength{\itemsep}{-0.05in}

\item
it obviously consists of roughly $4\times15$ the presently designed AMANDA
array, leading to an "effective" volume of $\sim6\times10^8$~m$^3$.
Importantly, the characteristics of the detector, including threshold in the
GeV-energy range, are the same as those of the AMANDA array module.

\item
the $1+7+7$ supermodule configuration, looked at as a whole, instruments a
1~km$^3$ cylinder with diameter and height of 1000~m with optical modules.
High-energy muons will be superbly reconstructed as they can produce triggers
in 2 or more of the supermodules spaced by
large distance.  Reaching more than one supermodule (range of 250~m) requires
muon energies in excess of 50~GeV. We note that this is the energy for which a
neutrino telescope has optimal sensitivity to a typical $E^{-2}$ source
(background falls with threshold energy, and until about 1~TeV little signal is
lost).

\end{enumerate}

\vspace*{-.5cm}

\begin{figure}[h]
\centering
\epsfxsize=7.5cm
\hspace{0in}\epsffile{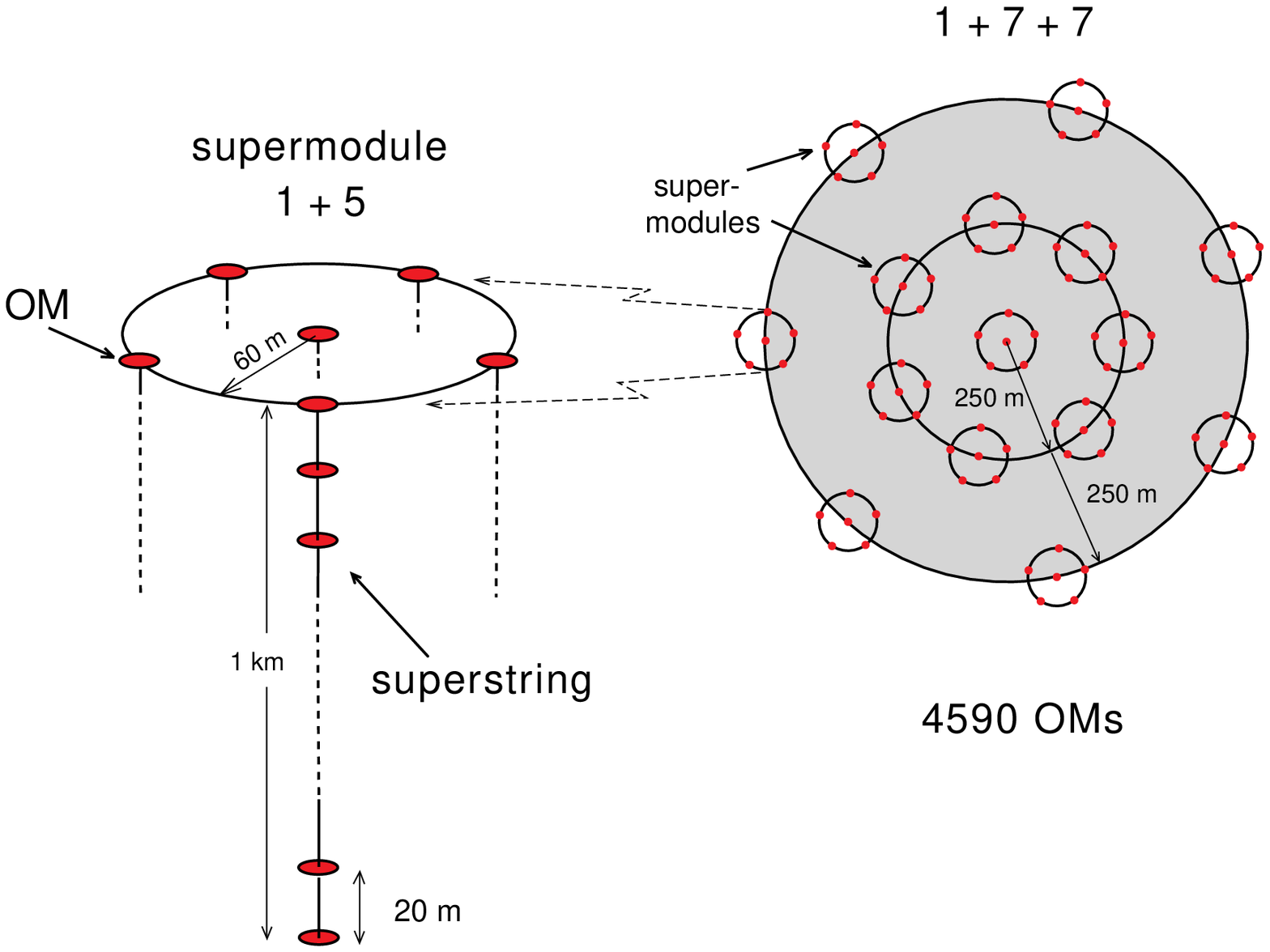}

\small Fig.~3
\end{figure}

\vspace*{-.2cm}

\noindent
Alternate methods to reach the 1~km scale have been discussed by Learned and
Roberts\cite{Roberts}.

What are the construction costs for such a detector? AMANDA's strings (with 10
OMs) cost \$150,000 including deployment. By naive scaling the final cost of
the postulated $1+7+7$ array of supermodules is of order \$50 million, still
below that of Superkamiokande (with $11{,}200 \times 20$~inch photomultiplier
tubes in a 40~m diameter by 40~m high stainless steel tank in a deep mine). It
is clear that the naive estimate makes several approximations over- and
underestimating the actual cost.

\section*{\uppercase{Acknowledgements}}

I would like to thank Tom Gaisser, Todor Stanev, John Jacobsen, Karl Mannheim
and Ricardo Vazquez for discussions.
This work was supported in part by the University of Wisconsin
Research Committee with funds granted by the Wisconsin Alumni Research
Foundation, and in part by the U.S.~Department of Energy under Contract
No.~DE-AC02-76ER00881.


\begin{thebibliography}{999}
%
\let\sl=\it
\bibitem{PR}
T.~K.~Gaisser, F.~Halzen and T.~Stanev, {\sl Physics Reports}, in press.

\bibitem{Venya}
V.S.~Berezinsky {\em et al.}, {\sl Proc.\ of the Astrophysics of Cosmic Rays};
Elsevier, New York (1991).

\bibitem{Mkr}
M.~Punch {\em et al.}, {\sl Nature} {\bf 358}, 477--478 (1992).

\bibitem{bier}
K. Mannheim and P.L. Biermann, {\it Astron. Astrophys.} {\bf 22}, 211 (1989);
K.~Mannheim and P.L. Biermann, {\it Astron. Astrophys.} {\bf 253}, L21 (1992);
K.~Mannheim, {\it Astron.\ Astrophys.} {\bf 269}, 67 (1993).

\bibitem{Ray}
A.P. Szabo and R.J. Protheroe, in {\it Proc.\ High Energy Neutrino Astrophysics
Workshop} (Univ.\ of Hawaii, March 1992, eds.\ V.J. Stenger, J.G. Learned,
S.~Pakvasa and X. Tata, World Scientific, Singapore).

\bibitem{Floyd}
F.W.~Stecker, C.~Done, M.H.~Salamon and P.~Sommers, {\it Phys.\ Rev.\ Lett.}
{\bf 66}, 2697 (1991) and {\bf 69}, 2738(E) (1992).

\bibitem{Super}
F. Halzen, J. E. Jacobsen and E. Zas, {\it Phys. Rev.} {\bf D49}, 1758 (1994).

\bibitem{SeckelDM}
J.~R.~Primack, B.~Sadoulet, and D.~Seckel, {\it Ann.\ Rev.\ Nucl.\ Part.\ Sci.}
{\bf B38}, 751 (1988).

\bibitem{Kamionkowski}
 F.~Halzen, M.~Kamionkowski, and T.~Stelzer, {\it Phys.\ Rev.} {\bf
D45}, 4439 (1992).

\bibitem{Learnedprime}
J.G.~Learned, {\it Proc.\ of the 13th European Cosmic Ray Symposium}, Geneva
1992, CERN (1992).

\bibitem{Roberts}
J.~G.~Learned and A.~Roberts, {\it Proceedings of the 23$^{rd}$ International
Cosmic Ray Conference}, Calgary, Canada  (1993);  F.~Halzen and J.G.~Learned,
{\it Proc.\ of the Fifth
International Symposium on Neutrino Telescopes}, Venice (1993),
ed.\ by  M.~Baldo-Ceolin.

\end{thebibliography}
\end{document}